\documentstyle[preprint,aps]{revtex}

\def\pr{{\em Phys. Rev., }}
\def\prb{{\em Phys. Rev., {\bf B}}}
\def\prl{{\em Phys. Rev. Lett., }}

\def\mplb{{\em Mod. Phys. Lett., \bf{ B}}}
\def\zpb{{\em Z. Phys., {\bf B}}}

\def\d{\partial}
\def\beq{\begin{equation}}
\def\eeq{\end{equation}}

\begin{document}

\draft
\title{Distribution of Wigner delay time from single channel disordered 
systems}
\author{Sandeep K. Joshi\cite{jos} and A. M. Jayannavar\cite{amj}}
\address{Institute of Physics, Sachivalaya Marg, Bhubaneswar 751 005, India}

\maketitle

\begin{abstract}

We consider the scattering of an electron from a semi-infinite
one-dimensional random medium. The random medium is characterized by
force, $-\d V/\d L$ being the basic random variable. We obtain an
analytical expression for the stationary delay time ($\tau$) distribution
$P_s(\tau)$ within a random phase approximation. Our result agrees with
earlier analytical expressions, where the random potential is taken to be
of different kind, indicating universality of the delay time distribution,
i.e., delay time distribution is independent of the nature of disorder.

\pacs{{\bf Keywords}: A. disordered systems, D. electronic transport, D.
quantum localization}
\end{abstract}

In recent years universal parametric correlations of phase shifts and
delay times in mesoscopic systems is being studied intensively\cite{yan}. 
The time delay in a scattering event (or duration of a collision event) is
an interesting aspect in itself in the general theory of quantum
scattering. Wigner was the first to establish the relation between the
time delay and the energy derivative of scattering phase
shift\cite{wigner}. Distribution of delay times in quantum chaotic regime
have been shown to be universal as it depends only on the symmetry
property of the Hamiltonian or scattering matrix\cite{yan,gopar}. The
delay time statistics is intimately connected with the issue of dynamic
admittance of microstructres (or mesoscopic systems)\cite{gopar,thomas},
for example quantum capacitance and its fluctuation\cite{nku}. The wave
packet incident on the surface a sample is not backscattered (or
reflected) immediately.  There will be some time delay before it is
reflected. This leads to a non-cancellation of the instantaneous currents
at the surface involving the incident and the reflected wave. This in-turn
is expected to lead to a low temperature $1/f$-type noise for the
fluctuating surface currents in the random
systems\cite{pendry,jayan,hein}. The study of change of density of states
due to scatterer is also directly related to the phase derivative of
scattering phase shift with respect to the energy, i.e., to the delay
time.

The distribution of delay time and its correlations in higher dimensions,
where system exhibits the Anderson localization has not been addressed so
far.  The first study of the stationary distribution $P_s(\tau)$ of delay
time $\tau$ for a disordered semi-infinite one-dimensional chain was
carried out in reference \cite{jayan}. Here authors used the invariant
imbedding approach. The underlying random potential $V(x)$ is treated as a
Gaussian white noise with zero mean. Using the random phase approximation
(RPA)  analytical expression for the $P_s(\tau)$ was obtained, which
exhibits $1/\tau^2$ dependence for the tail of the delay distribution.
Further developments for $P_s(\tau)$ using supersymmetric potentials lead
to same distribution function $P_s(\tau)$ within RPA. This has lead to a
conjecture that within RPA, $P_s(\tau)$ is independent of nature of
disorder and, hence, is universal\cite{comtet}. Our recent numerical study
has clearly indicated that long time delay distribution is universal
beyond RPA\cite{joshi}. In our present work we calculate the distribution
of delay time where we take $-\d V/\d x$ as the basic random variable with
delta correlated Gaussian distribution and we obtain analytical expression
for $P_s(\tau)$ in RPA. The stationary distribution has the same
functional form obtained earlier with different random potential
indicating the universal nature of $P_s(\tau)$. 

The model Hamiltonian for the 1-D disordered system is
\beq
\label{scheqn}
H~=~\frac{-\hbar^2}{2m} \frac{\d^2~~}{\d x^2} + V(x)
\eeq
where $V(x)$ for $0<x<L$ is the random potential. The disordered sample
extends from $x=0$ to $X=L$, the two ends being connected to perfect
leads. Consider an electron of wave number $k$ incident at $x=L$ from
right. It is partially reflected with the complex reflection coefficient
$R(L)$ and partially transmitted. The transmission and reflection
coefficients are emergent quantities of direct physical interest for the
conductance problem. The method of invariant imbedding was proposed
originally by S. Chandrashekhar in the context of radiative transfer
through stellar atmosphere\cite{chand}. His method consists of viewing the
given sample of length $L$ as imbedded in a larger sample of length
$L+\Delta L$ and then setting up an equation for the resulting change in
the S-matrix $\Delta S$ as $\Delta L \rightarrow 0$. In order to look for
the complex reflection coefficient, we transform Eqn. \ref{scheqn} to the
invariant imbedding equation for the complex reflection amplitude
$R(L)=|R(L)| Exp(i \theta(L))$. The evolution equation for $R(L)$ is now
given by \cite{bellwing} 
\beq 
\label{invimeq} 
\frac{\d R(L)}{\d L}=f_1(L)  + 2if_0(L) R(L) - f_1(L) R^2(L),
\eeq 

with $$f_1(L)=\frac{2}{k(L)}\frac{\d k}{\d L},$$ $$f_0(L)=k(L)$$ and $$
k^2 = \frac{2m}{\hbar^2} \left ( E - V(L) \right ). $$
	
\noindent The above equation Eqn. \ref{invimeq} was first studied in Ref. 
\cite{nku2} to evaluate the resistance and its fluctuation in a disordered
quantum wire. The invariant imbedding method has been generalized to
N-channel case and in an equivalent random phase approximation has lead to
DMPK (Dorokhov-Mello-Pereyra-Kumar) equation\cite{dmpk}, using which
coherent transport properties have been analyzed extensively in mesoscopic
systems. 

In the present problem we consider random potential $V(L)$ to be
bounded having a small amplitude. However, $-\d V/\d L$ can be unbounded
and we treat $\xi(L)=-\d V/\d L$ as our basic random variable. The
energy of incident electron is assumed to be large, i.e., much larger than
the magnitude of the upper bound on the potential. In that case we have

\beq
\label{f0f1}
f_1(L) =  \frac{-1}{E} \frac{\d V(L)}{\d L} 
~~\mathrm{and}~~
f_0(L) = \sqrt{\frac{2m}{\hbar^2}} \sqrt{E}
\eeq

We take $\xi(L)$ to be Gaussian delta correlated random number with zero
mean and

\beq
\label{dsalpha}
\left < \xi(L) \xi(L^\prime) \right > = 2\alpha~\delta(L-L^\prime).
\eeq

Here, the $<....>$ denotes the ensemble average with respect to the
realizations of the stochastic variable $\xi$ and $\alpha$ denotes the
strength of the disorder. The equation for the phase ($\theta$) is readily
obtained from Eqn. \ref{invimeq} as

\beq
\label{thetaeqn}
\frac{\d \theta}{\d L} = 2 \sqrt{\frac{2m}{\hbar^2}} \sqrt{E} - 2
\frac{\xi(L)}{E} sin(\theta)
\eeq

where we have set $|R|=1$ since we will be interested in the limit
$L\rightarrow\infty$ (semi-infinite medium), i.e., total back-reflection
with probability one. The delay time is given by $\tau=\hbar\d \theta/\d
E$. Differentiation of Eqn. \ref{thetaeqn} with respect to $E$ leads to
the following equation for the evolution of $\tau$:

\beq
\label{taueqn}
\frac{\d \tau}{\d L} = \frac{\sqrt{2m}}{\sqrt{E}} +
\frac{2\hbar}{E^2}\xi(L) sin(\theta) - \frac{2}{E} \xi(L) cos(\theta) \tau
\eeq

From Eqns. \ref{thetaeqn} and \ref{taueqn} one can obtain readily obtain
the equation governing the evolution of the joint probability distribution
$W(\tau,\theta;L)$ for $\theta$ and $\tau$ by using the Van Kampen
lemma\cite{vankam} and Novikov's theorem\cite{novikov,nov1,nov2}. In our
case, however, we are interested only in the marginal probability
distribution $P(\tau;L)= \int_0^{2\pi} W(\tau,\theta;L) d\theta$ of delay
time $\tau$. The delay time being the derivative of phase we expect it to
fluctuate much more rapidly as compared to the phase itself. We therefore
make the decoupling approximation, as done in earlier literature by
\cite{jayan,hein}, treating $\theta$ and $\tau$ as statistically
independent variables in the large length ($L$)  limit. As mentioned
earlier, we are interested in the case of high energy particles ($E \gg
V$) and in this limit the distribution of $\theta$ becomes
uniform\cite{stone,jayanpra,jayanssc}, i.e., $P(\theta)=1/2\pi$. This is
generally referred to as random phase approximation (RPA). Within the
above mentioned approximations after a straight forward algebra the
evolution equation for $P(\tau;L)$ becomes

\beq 
\label{dpdleqn} 
\frac{\d P}{\d L}=\hbar \frac{\d}{\d \tau} \left
\{\frac{2\alpha\hbar}{E^4} \frac{\d P}{\d \tau} -
\frac{\sqrt{2m}}{\sqrt{E}} P + \frac{4\alpha}{\hbar E^2} \tau P +
\frac{2\alpha}{\hbar E^2} \tau^2 \frac{\d P}{\d \tau} \right \}
\eeq

The stationary distribution $P_s(\tau)$ for $\tau$ in the limit $L
\rightarrow \infty$ can be obtained by setting $\d P/\d L =0$. We get the
following expression for normalized $P_s(\tau)$

\beq
\label{pst}
P_s(\tau) = \frac{\lambda e^{\lambda tan^{-1}\tau}}{(e^{\lambda \pi/2}
-1)(1 + \tau^2)}
\eeq

In the above expression we have redefined $\tau$ in a dimensionless form
$\tau \equiv \tau E/\hbar$ and $\lambda = \sqrt{2mE}~E^2/(2\alpha\hbar)$. 
The most probable value of $\tau$ occurs at $\tau=\lambda/2$. As $\tau
\rightarrow \infty$, $P_s(\tau) \rightarrow 1/\tau^2$, i.e., the
distribution has a long time tail which goes as $1/\tau^2$. This leads to
the logarithmic divergence of the average value of $\tau$ indicating that
the origin of such a tail is due, presumably, to the Azbel
resonances\cite{azbel} which make Landauer's four probe conductance
infinite even for a finite sample. In case of these resonant
realizations,the time spent by the particle inside the medium is large as
it travels a large distance before getting reflected. It is now well
established that coherent interference effects, due to elastic scattering
by the serial static disorder lead to localization of eigenstates for
arbitrary weak disorder. The localization length $l$ of these eigenstates
is a self averaging quantity\cite{nku3} and in a one-dimensional system it
is directly proportional to elastic mean free path. The most probable
value $\tau_{max}$ of $\tau$ is proportional to a time taken by a particle
to traverse a distance of the order of localization length $l$,
$\tau_{max} \propto 2l/(\hbar k/m)$, where k is the incident energy. From
this one can readily obtain the behavior of localization length on the
material parameters, namely, $l \propto E^2/\alpha$.

Our above analytical expression has the same functional form as obtained
earlier where potential itself is treated as a Gaussian random variable
using a different invariant imbedding equation. From this we conclude that
two different models of random variable lead to the same universal
distribution of the delay time. Thus reinforcing the conjecture of
universal behavior of the delay time distribution, independent of nature
of disorder within RPA.

\end{document}